\documentclass[12pt]{article}
\usepackage{amsmath}
\usepackage{graphicx}
\usepackage{amsfonts}
\usepackage{amssymb}

\begin{document}

\def\half{{\textstyle{1\over2}}}
\let\a=\alpha \let\b=\beta \let\g=\gamma \let\d=\delta \let\e=\epsilon
\let\z=\zeta \let\h=\eta \let\th=\theta \let\i=\iota \let\k=\kappa
\let\l=\lambda \let\m=\mu \let\n=\nu \let\x=\xi \let\p=\pi \let\r=\rho
\let\s=\sigma \let\t=\tau \let\u=\upsilon \let\f=\phi \let\c=\chi \let\y=\psi
\let\vp=\varphi \let\vep=\varepsilon
\let\w=\omega      \let\G=\Gamma \let\D=\Delta \let\Th=\Theta \let\L=\Lambda
\let\X=\Xi \let\P=\Pi \let\S=\Sigma \let\U=\Upsilon \let\Y=\Psi
\let\C=\Chi \let\W=\Omega
\let\la=\label \let\ci=\cite \let\re=\ref
\let\se=\section \let\sse=\subsection \let\ssse=\subsubsection
\def\nn{\nonumber} \def\bd{\begin{document}} \def\ed{\end{document}}
\def\ds{\documentstyle} \let\fr=\frac \let\bl=\bigl \let\br=\bigr
\let\Br=\Bigr \let\Bl=\Bigl
\let\bm=\bibitem
\let\na=\nabla
\def\tU{{\widetilde U}}
\let\pa=\partial \let\ov=\overline
\def\ie{{\it i.e.\ }}
\newcommand{\be}{\begin{equation}}
\newcommand{\ee}{\end{equation}}
\def\ba{\begin{array}}
\def\ea{\end{array}}
\def\ft#1#2{{\textstyle{{\scriptstyle #1}\over {\scriptstyle #2}}}}
\def\fft#1#2{{#1 \over #2}}
\def\F#1#2{{ F_{#1}^{(#2)} }}
\def\cF#1#2{{ {\cal F}_{#1}^{(#2)} }}

\def\R{{\bf R}}
\def\sst#1{{\scriptscriptstyle #1}}
\def\oneone{\rlap 1\mkern4mu{\rm l}}
\def\e7{E_{7(+7)}}
\def\td{\tilde}
\def\wtd{\widetilde}
\def\im{{\rm i}}
\def\bog{Bogomol'nyi\ }
\newcommand{\ho}[1]{$\, ^{#1}$}
\newcommand{\hoch}[1]{$\, ^{#1}$}
\newcommand{\bea}{\begin{eqnarray}}
\newcommand{\eea}{\end{eqnarray}}
\newcommand{\ra}{\rightarrow}
\newcommand{\lra}{\longrightarrow}
\newcommand{\Lra}{\Leftrightarrow}
\newcommand{\ap}{\alpha^\prime}
\newcommand{\bp}{\tilde \beta^\prime}
\newcommand{\cB}{{\cal B}}
\newcommand{\cO}{{\cal O}}
\newcommand{\vecx}{\vec{x}}
\newcommand{\vecy}{\vec{y}}
\newcommand{\vecp}{\vec{p}}
\newcommand{\vecq}{\vec{q}}
\newcommand{\tr}{{\rm tr} }
\newcommand{\Tr}{{\rm Tr} }
\newcommand{\NP}{Nucl. Phys. }

\newcommand{\cL}{{\cal L}}
\newcommand{\cA}{{\cal A}}
\newcommand{\cD}{{\cal D}}

\def\ve{\varepsilon}
\def\vf{\varphi}
\def\F{\Phi}
\def\wg{\wedge}

\def\ve{\varepsilon}
\def\vf{\varphi}
\def\F{\Phi}
\def\wg{\wedge}

\newcommand{\auth}{AUTHORS}

\def\thb{\bar{\theta}}
\def\Thb{\bar{\Theta}}
\def\barp{\bar{p}}
\def\barq{\bar{q}}
\def\barc{\bar{c}}
\def\bard{\bar{d}}
\def\e{\epsilon}

\def\th{\theta}
\def\Th{\Theta}
\def\vth{\vartheta}
\def\btheta{{\bar\theta}}
\def\ttheta{{{\tilde\theta}}}
\def\bttheta{{{\bar\ttheta}}}
\def\vth{\vartheta}

\def\ra{\rightarrow}
\def\N{{\cal N}}
\def\F{{\cal F}}
\def\uM{\underline{M}}
\def\uN{\underline{N}}
\def\uP{\underline{P}}
\def\cc{\circ}
\def\eqv{\equiv}

\def\ni{\noindent}

\def\Ep{E^{{}^{(+)}}}
\def\Em{E^{{}^{(-)}}}

\def\Mp{M^{{}^{(+)}}}
\def\Mm{M^{{}^{(-)}}}

\def\xb{\bar{x}}
\def\xib{\bar{\xi}}
\def\lb{\bar{\l}}
\def\tb{\bar{t}}
\def\vb{\bar{v}}

\def\an{|n|}
\def\xt{\tilde{x}}
\def\pnr{(p_n^{(r)}}
\def\Xd{\dot{X}}
\def\amn{{\a_{-n}}}
\def\At{\tilde{A}}
\def\Bt{\tilde{B}}
\def\ola{\overleftarrow}
\def\ora{\overrightarrow}
\def\at{\tilde{\a}}
\def\st{\star}
\def\qb{\bar{q}}
\def\qt{\tilde{q}}

\overfullrule=0pt
\parskip=2pt
\parindent=12pt
\headheight=0in \headsep=0in \topmargin=0in
\oddsidemargin=0in

\begin{flushleft}
\hfill hep-th/0403218 \\
OHSTPY-HEP-T-04-004
\bigskip
\end{flushleft}

\bigskip
\thispagestyle{empty}

\vspace{1cm}
\begin{center}
{\Large\bf  Absorption of a Quantum in a D1/D5 System
             }

\vspace{1cm}

\vspace{.5cm}
{I.Y. Park}\\
{\em Department of Physics, The Ohio State University,\\
 Columbus, OH 43210, USA
\bigskip}

\end{center}

 \vspace{1.3cm}
 \begin{abstract}
 In {\em Nucl.\ Phys.}\ B {\bf 615}, 285 (2001) [arXiv:hep-th/0107113],
the wave equation for a minimally coupled scalar was studied in
the geometry of a D1-D5 system with non-zero angular momentum.
The probability for a quantum to enter the throat was computed by
taking small a parameter $\g$ which is associated with the value
of the angular momentum. In the leading order in $\g$, the result
was found to agree with the dual CFT result. In this note, we
 report on an observation that there are corrections of higher order
 in $\g$. Our results should be useful for determining the higher order
 correction terms that the dual CFT  needs in order to incorporate the
 presence of the `capped' geometry.
 \end{abstract}

\newpage

A D1/D5 brane system has been a valuable asset for studying black
hole physics leading to many fruitful results in the calculations
of entropy and radiation rate. For this system, the microscopic
consideration yielded exactly the same value as the Bekenstein
entropy \cite{sv,cm}. Similarly precise agreement was found
between its low energy radiation rate and the Hawking radiation
of the corresponding black hole \cite{dm,ms}.

It has also served as a stage where one tackles the black hole
information puzzle. A recent attempt to resolve the information
puzzle can be found in \cite{Mathur:2002ie,Lunin:2001jy}. The
system that was considered in these works is not the conventional
D1/D5 brane which has an infinite throat, but rather the one with
its throat `capped' off at a finite distance in the conical
singularity. The way the throat caps off maps to a dual CFT
microstate.  The mapping was established in
\cite{Lunin:2001jy,Lunin:2002bj} for special family of metrics
providing another test for AdS/CFT correspondence.

In \cite{Lunin:2001dt}, the wave equation for a minimally coupled
scalar was studied in the geometry of these special family of
metrics. The truncated throat was called a `tube'.
 Most of the low energy quanta that
are sent toward the throat from spatial infinity get reflected
back to the infinity when they arrive at the start of the throat.
But there is a small probability for them to enter the throat and
travel until they reach the end of the throat. Once reaching the
end, they get reflected back to the start of the throat where they
have the same small probability to exit from the tube.

 The probability for a quantum to enter the throat was obtained by
 the same authors. They also computed the `time delay' for two low frequency
  quanta that emerge from the tube successively.
 However for both of these computations a parameter $\g$
 which is associated with the value of the angular momentum was
 taken small. In the leading order in $\g$, agreements were
 found with the corresponding CFT computations.
  Since the presence of $\g$ itself is tied with the fact that the
  geometry is capped, it is worth re-analyzing the problem
  in the sub-leading order in $\g$ in order to see
  whether corrections occur.

 This is, in fact, what we will do in this note: we keep $\g$
 arbitrary, and re-compute the time delay and the probability.
 Also in \cite{Lunin:2001dt} the computation was done
 with null momentum along the compact direction of D1 brane
 wrapping. We will lift this condition too since it is not difficult to
 do so. As we will see below, the time delay turns out to be the same.
 However, the probability gets corrected in $\g$ (and in $\l$ as well).
 In other words, for a generic value of
 the parameter $\g$ the gravity calculation of the probability gives a different
 result than the one obtained from the CFT analysis although the former gets
 reduced to the latter in the small $\g$ limit. The discrepancy could probably be
 accounted by putting correction terms, which arise from a DBI action, in
 the CFT to incorporate the presence of the `capped' geometry.
 Our results should be useful to determine them.

\bigskip

We begin by reviewing \cite{Lunin:2001dt}. Let $R$ be the radius
of the circle that is common to the D1 and D5 branes, and $n_1,
n_5$ the number of D1, D5 branes respectively. The metric for the
rotating D1/D5 brane system \cite{cvetic, mm} is:
 \bea\label{MaldToCompare} d{s}^2&=&-\frac{1}{h}(d{t}^2-d{ y}^2)+
hf\left(d\theta^2+\frac{d{r}^2}{{r}^2+a^2}\right)
-\frac{2a\sqrt{Q_1 Q_5}}{hf}\left(\cos^2\theta d{ y}d\psi+
\sin^2\theta d{ t}d\phi\right)\nonumber\\
&+&h\left[ \left({
r}^2+\frac{a^2Q_1Q_5\cos^2\theta}{h^2f^2}\right) \cos^2\theta
d\psi^2+ \left({
r}^2+a^2-\frac{a^2Q_1Q_5\sin^2\theta}{h^2f^2}\right)
\sin^2\theta d\phi^2\right],\nonumber \\
 \eea
where
\be\label{defFHProp} f={ r}^2+a^2\cos^2\theta,\qquad
h=\left[\left(1+\frac{Q_1}{f}\right)\left(1+\frac{Q_5}{f}\right)\right]^{1/2}
 \ee
The parameter $a$ has a physical meaning that the throat ends at
$r \sim a$. It was re-written in \cite{Lunin:2001dt} in terms of
a dimensionless parameter $\gamma=\frac{2J}{n_1n_5}$ where $J$
represents the angular momentum:
 \be a=\frac{\sqrt{Q_1Q_5}}{R}\gamma \label{a}
 \ee
We take $R\gg (Q_1Q_5)^{1/4}$. Consider the wave equation for a
minimally coupled scalar in this background. For a massless
scalar, it is
\be\label{GenScalEqn} \Box\Phi=\frac{1}{\sqrt{-g}}\pa_\mu
\left(\sqrt{-g}g^{\mu\nu}\pa_\nu\Phi\right)=0.
 \ee
Writing the solution  in the following form:
 \be \Phi(t,r,\theta,\phi,\psi,y)=\exp(-i\tilde\omega
 t+im\phi+in\psi+i\tilde\l y) {\tilde\Phi}(r,\theta).
 \ee
with
 \be
\tilde\Phi(r,\theta)=H(r)\Theta(\theta),
 \ee
equation (\ref{GenScalEqn}) turns into a set of two equations, one
for the
 radial part and the other for the angular part. The existence of separation
of variables for such a wave equation was shown in
\cite{larsentwo}. Denoting the eigenvalue of the angular equation
by $\L$, its approximate value was determined in
\cite{Lunin:2001dt}:
 \be\label{EigenLaAppr}
 \Lambda=l(l+2)+O((a\tilde\omega)^2)+O((a\tilde\lambda)^2)
 \ee
where $l$ is a nonnegative integer that specifies the spherical
harmonic of the angular solution. By introducing a new coordinate
\be x=\frac{r^2 R^2}{Q_1Q_5} \ee and dimensionless parameters
$\omega=R\tilde\omega$, $\l=R\tilde\l$ the radial equation can be
put into the following form,
 \bea\label{RadEqn}
&&4\frac{d}{dx}\left(x(x+\gamma^2)\frac{dH}{dx}\right)+
\left\{(\omega^2-\l^2)\left[\frac{Q_1Q_5}{R^4}x+\frac{Q_1+Q_5}{R^2}\right]
\right.\nonumber\\
&&\qquad\left.+\frac{(\omega-m\gamma)^2}{x+\gamma^2}
-\frac{(\l+n\gamma)^2}{x} \right\}H-\Lambda H=0
 \eea
A solution was found in \cite{Lunin:2001dt} imposing
 $\lambda=0$ and taking $\g\rightarrow 0$. Here we relax these constraints.

\bigskip

As in \cite{Lunin:2001dt}, we solve (\ref{RadEqn}) in two
asymptotic regions and match the solutions obtained in each
region. First consider the large-$x$ region. In this region we can
borrow the corresponding steps of \cite{Lunin:2001dt} with minimal
modifications. For self-containedness of this note, we enclose
some details: (\ref{RadEqn}) simplifies to
 \be 4x^2H''+8xH'+(\omega^2-\l^2)
 \left\{\frac{Q_1Q_5}{R^4}x+\frac{Q_1+Q_5}{R^2}\right\}H-\Lambda
 H=0
 \ee
which is the same as the corresponding equation in
\cite{Lunin:2001dt} other than the shift, $w^2\rightarrow
\omega^2-\l^2$. The general solution is a linear combination of
Bessel functions, but now with a shifted argument,
 \be\label{OuterHyperAc} H_{out}(x)=\frac{1}{\sqrt{x}}\left[C_1
 J_\nu\left(\sqrt{\frac{Q_1Q_5(\omega^2-\l^2)x }{R^4}}\right)+ C_2
 J_{-\nu}\left(\sqrt{\frac{Q_1Q_5(\omega^2-\l^2)x }{R^4}}\right)\right],
 \ee
where
 \be\label{DefNu}
\nu=\left(1+\Lambda-(\omega^2-\l^2)\frac{Q_1+Q_5}{R^2}\right)^{1/2}\equiv
l+1+\e
 \ee
 and
 \be \e\approx
-\frac{(\omega^2-\l^2)}{l+1}\frac{Q_1+Q_5}{2R^2} \label{eone}
 \ee
Expanding the Bessel functions in the leading order in the region
where
 \be
  \frac{Q_1Q_5 (\omega^2-\l^2)x }{R^4}\ll 1
 \ee
one gets
 \bea\label{AsympOut}
  H_{out}(x) \approx && \frac{1}{\sqrt{x}}
\left(\frac{Q_1Q_5(\omega^2-\l^2)x
}{4R^4}\right)^{-\frac{l+1}{2}} \nn\\
 && \left[-(-1)^l C_2\e l!+
\frac{C_1-(-1)^l C_2}{(l+1)!}\left(\frac{Q_1Q_5(\omega^2-\l^2)x
}{4R^4}\right)^{l+1} \right].
 \eea
\\
Now let's turn to the inner region. The inner region is defined by
 \be
 x\ll \frac{(Q_1+Q_5)R^2}{Q_1Q_5}
 \ee
Then (\ref{RadEqn}) can be approximated to:
 \bea\label{RadEqnPr}
 &&4\frac{d}{dx}\left(x(x+\gamma^2)\frac{dH}{dx}\right)\!+
 \!\left\{(\omega^2-\l^2)\left[\frac{Q_1+Q_5}{R^2}\right]\!\!
 +\!\!\frac{(\omega-m\gamma)^2}{x+\gamma^2}\! -\!\frac{(\l+n\gamma)^2}{x}
 \right\}H\!\!=\Lambda H\!\! \nn\\
 \eea
By taking
 \be
 H_{in}(x)=x^\alpha(\gamma^2+x)^\beta G(x),
 \ee
where
 \be
 \a=\fr12 \left(n+\fr{\l}{\g}\right) \;\;,\;\;
 \b=\fr{w-\g m}{2\g}
 \ee
equation (\ref{RadEqnPr}) is converted to the hypergeometric
differential equation
 \bea\label{radialEqnGamma}
&&4x(x+\gamma^2)G''+4\left[2x(\alpha+\beta+1)+\gamma^2(1+2\alpha)\right]
G'\nonumber\\
&&+\left\{4(\alpha+\beta)(\alpha+\beta+1)+
(\omega^2-\l^2)\frac{Q_1+Q_5}{R^2}-\Lambda \right\}G=0
 \eea
The solution that is regular at $x=0$ is
 \bea\label{SolveHyperG} G(x)&=&F\left(p,q;
1 + 2\alpha;-\frac{x}{\gamma^2}\right)
 \eea
with
 \bea
p&=&\frac{1}{2}+\alpha+\beta +
\frac{1}{2}\sqrt{1+\Lambda-(\omega^2-\l^2)\frac{Q_1+Q_5}{R^2}}\\
q&=&\frac{1}{2}+\alpha+\beta -
\frac{1}{2}\sqrt{1+\Lambda-(\omega^2-\l^2)\frac{Q_1+Q_5}{R^2}}
 \eea
For further manipulation, it is convenient to re-express $G(x)$
using one of the hypergeometric identities,
\bea\label{ExactG} G(x)&=&\frac{\Gamma(1 +
2\alpha)\Gamma(-{\nu'})}{
\Gamma(\frac{1}{2}+\alpha+\beta-\frac{{\nu'}}{2})
\Gamma(\frac{1}{2}+\alpha-\beta-\frac{{\nu'}}{2})}
\left(\frac{x}{\gamma^2}\right)^{-p}
F\left(p,p-2\alpha;{\nu'}+1;-\frac{\gamma^2}{x}\right)\nonumber\\
&+&\frac{\Gamma(1 + 2\alpha)\Gamma({\nu'})}{
\Gamma(\frac{1}{2}+\alpha+\beta+\frac{{\nu'}}{2})
\Gamma(\frac{1}{2}+\alpha-\beta+\frac{{\nu'}}{2})}
\left(\frac{x}{\gamma^2}\right)^{-q}
F\left(q,q-2\alpha;-{\nu'}+1;-\frac{\gamma^2}{x}\right)\nonumber
\\
\eea
In \cite{Lunin:2001dt}, the leading behavior of $G(x)$ was
obtained by first approximating the hypergeometric functions to
the Bessel functions. Here we do not take these intermediate
steps but rather directly deal with the hypergeometric functions.

Noting that the hypergeometric function has the following series
expansion
 \be F(p,p-2\alpha;c;-z)=\sum_{n=0}^\infty
\frac{\Gamma(p+n)}{\Gamma(p)}
\frac{\Gamma(p-2\alpha+n)}{\Gamma(p-2\alpha)}\frac{\Gamma(c)}{\Gamma(c+n)}
\frac{(-z)^n}{n!}
 \ee
one can approximate (\ref{ExactG}):
 \bea
 G(x)&\simeq &
 \fr{\G(1+2\a)\G(-\n\,')}{\G(\fr12+\a+\b-\fr{\n'}{2})\G(\fr12+\a-\b-\fr{\n'}{2})}
 \left(\fr{x}{\g^2}\right)^{-p} \nn\\
   &&+\;\; \fr{\G(1+2\a)\G(\n\,')}{\G(\fr12+\a+\b+\fr{\n'}{2})\G(\fr12+\a-\b+\fr{\n'}{2})}
 \left(\fr{x}{\g^2}\right)^{-q}(1+Y)
 \eea
where\footnote{In \cite{Lunin:2001dt}, another symbol $\e'$ was
introduced for the inner region: it is the same as $\e$. We simply
adopt their notation. }
 \bea
 Y & \equiv & \xi \;\fr{\G(-l-\e')}{\G(1-\e')}\fr{(-z)^{l+1}}{(l+1)!}\nn\\
   & \simeq &  \xi\;\fr{1}{\e'}\fr{z^{l+1}}{l!(l+1)!}\fr{1}{\G(1-\e')}
 \eea
with $z \equiv \fr{\g^2}{x}$ and
 \bea
  \xi &\equiv & \fr{\G(q+l+1)}{\G(q)}\fr{\G(q-2\a+l+1)}{\G(q-2\a)}
  \nn\\
   &=& (q+l)\cdots q \;(q-2\a+l)\cdots (q-2\a)
 \eea
Assuming $x \gg \g^2$
 \bea\label{AsympIn}
H_{in} & \simeq & x^{\a+\b}\;G(x) \nn\\
       & \simeq & \fr{\G(1+2\a)\G(-\n')}{\G(\fr12+\a+\b-\fr{\n'}{2})
       \G(\fr12+\a-\b-\fr{\n'}{2})}
 \left(\fr{x}{\g^2}\right)^{\a+\b-p}\g^{2(\a+\b)} \nn\\
   &&+\;\; \fr{\G(1+2\a)\G(\n')}{\G(\fr12+\a+\b+\fr{\n'}{2})\G(\fr12+\a-\b+\fr{\n'}{2})}
 \left(\fr{x}{\g^2}\right)^{\a+\b-q} \g^{2(\a+\b)}(1+Y) \nn\\
  & \simeq &\fr{\g^{1+2(\a+\b)}}{\sqrt{x}} \fr{\G(1+2\a)}{\G(\fr12+\a+\b+\fr{\n'}{2})
   \G(\fr12+\a-\b+\fr{\n'}{2})} \nn\\
  && \left[\;\; \fr{\G(\fr12+\a+\b+\fr{\n'}{2})
   \G(\fr12+\a-\b+\fr{\n'}{2})}
   {\G(\fr12+\a+\b-\fr{\n'}{2})\G(\fr12+\a-\b-\fr{\n'}{2})}
   \G(-\n')\left(\fr{x}{\g^2}\right)^{-\fr12(l+1)} \right. \nn\\
  && \left. +\fr{\G(\n')}{\G(1-\e')}\;\xi
  \fr{1}{\e'}\fr{1}{l!(l+1)!} \left(\fr{x}{\g^2}\right)^{-\fr12(l+1)}
  +\G(\n')\left(\fr{x}{\g^2}\right)^{\fr12(l+1)}\;\;\right]
 \eea
Imagine a low frequency quantum that is sent toward $r=0$ from
spatial infinity. As shown in \cite{Lunin:2001dt}, it is likely to
be reflected at the start of the throat. But there is a small
probability, $P$, for the quantum
 to enter the throat. Once it enters the throat, it travels to the
end of the throat. There it gets reflected back to the start of
the throat and could leave the throat toward spatial infinity
with the same probability, $P$. Consider a quantity $R$, whose
absolute square is the reflection coefficient. Its explicit
expression is obtained in the first equation of (4.22) of
\cite{Lunin:2001dt}, which we quote here,
 \bea
 R&=&e^{-i\pi\e}-(1-e^{-2i\pi\e})\frac{C_2(-1)^l}
 {C_1-C_2(-1)^l e^{-i\pi\e}}\nonumber\\ \label{R}
 \eea
Note that
 \bea
 \fr{C_2(-1)^l}{C_1-(-1)^lC_2\;e^{-i\pi\e}}
 & \simeq & \fr{C_2(-1)^l}{C_1-(-1)^lC_2\;(1-i\pi\e)} \nn\\
 & \simeq & \fr{C_2(-1)^l}{C_1-(-1)^lC_2 +(-1)^lC_2 i\pi\e} \nn\\
 &=& \fr{\fr{C_2(-1)^l}{C_1-(-1)^lC_2}}{1 +\fr{C_2(-1)^l}{C_1-(-1)^lC_2}
 i\pi\e}
 \eea
As we will see below, $\fr{C_2(-1)^l}{C_1-(-1)^lC_2}
 i\pi\e$ is small, therefore we have
  \bea
 \fr{C_2(-1)^l}{C_1-(-1)^lC_2\;e^{-i\pi\e}}
 & \simeq & \fr{C_2(-1)^l}{C_1-(-1)^lC_2} \label{csimple}
 \eea
To obtain the matching conditions, compare the inner solution
(\ref{AsympIn}) with the outer solution (\ref{AsympOut}).
Comparison of the coefficients of the negative power of $x$ gives
 \bea
 &&\left(\fr{Q_1Q_5(w^2-\l^2)}{4R^4}\right)^{-\fr{l+1}{2}}(-1)^{l+1}\;C_2 \;\e
 l!\nn\\
  \simeq  && {\g^{1+2(\a+\b)}}
\fr{\G(1+2\a)}{\G(\fr12+\a+\b+\fr{\n'}{2})
   \G(\fr12+\a-\b+\fr{\n'}{2})} \nn\\
  && \left[\;\; \fr{\G(\fr12+\a+\b+\fr{\n'}{2})
   \G(\fr12+\a-\b+\fr{\n'}{2})}
   {\G(\fr12+\a+\b-\fr{\n'}{2})\G(\fr12+\a-\b-\fr{\n'}{2})}
   \G(-\n')\;\g^{l+1} \right. \nn\\
  && \left. \;\;+\;\fr{\G(\n')}{\G(1-\e')}\;\xi
  \fr{1}{\e'}\fr{1}{l!(l+1)!}\;\g^{l+1}
  \;\;\right]
 \eea
Solving for $C_2$, one gets
 \bea
 C_2
 & \simeq & \left(\fr{Q_1Q_5(w^2-\l^2)}{4R^4}\right)^{\fr{l+1}{2}}
 \fr{(-1)^{l+1}
 {\g^{1+2(\a+\b)+l+1}}\G(1+2\a)}{\G(\fr12+\a+\b+\fr{\n'}{2})
   \G(\fr12+\a-\b+\fr{\n'}{2})} \nn\\
  && \fr{1}{\e\e'l! (l+1)!}\left[\; \fr{\G(\fr12+\a+\b+\fr{\n'}{2})
   \G(\fr12+\a-\b+\fr{\n'}{2})}
   {\G(\fr12+\a+\b-\fr{\n'}{2})\G(\fr12+\a-\b-\fr{\n'}{2})}\;(-1)^l
  +\fr{\G(\n')}{\G(1-\e')}\;\xi
  \fr{1}{l!}\;
  \right]\nn\\  \label{c2}
 \eea
The other matching condition is
 \bea
 && \left(\fr{Q_1Q_5(w^2-\l^2)}{4R^4}\right)^{\fr{l+1}{2}} \fr{C_1-(-1)^lC_2}{(l+1)!}
  \nn\\
  \simeq && {\g^{1+2(\a+\b)-l-1}}\;\fr{\G(1+2\a)\G(\n')}{\G(\fr12+\a+\b+\fr{\n'}{2})
   \G(\fr12+\a-\b+\fr{\n'}{2})}
 \eea
which gives
 \bea
 && \fr{1}{C_1-(-1)^l C_2} \\
 \simeq &&  \left(\fr{Q_1Q_5(w^2-\l^2)}{4R^4}\right)^{\fr{l+1}{2}}
 \fr{\g^{l-2(\a+\b)}}{(l+1)!\;\G(\n')}\fr{\G(\fr12+\a+\b+\fr{\n'}{2})
   \G(\fr12+\a-\b+\fr{\n'}{2})}{\G(1+2\a)} \nn
 \eea
From this and (\ref{c2}), one gets
 \bea
   \fr{C_2}{C_1-(-1)^lC_2}& \simeq &\left(\fr{Q_1Q_5(w^2-\l^2)}{4R^4}\right)^{{l+1}}
   (-1)^{l+1}\g^{2(l+1)} \fr{1}{\e\e'l![(l+1)!]^2}\fr{1}{\G(\n')}
   \\
   && \left[\;\; \fr{\G(\fr12+\a+\b+\fr{\n'}{2})
   \G(\fr12+\a-\b+\fr{\n'}{2})}
   {\G(\fr12+\a+\b-\fr{\n'}{2})\G(\fr12+\a-\b-\fr{\n'}{2})}\;(-1)^l
  \;\;+\;\fr{\G(\n')}{\G(1-\e')}\;\xi
  \fr{1}{l!}\; \nn
  \;\;\right]
 \eea
Noting that
 \bea
 \G(\n')&=& \G(l+1+\e')\simeq l!
  \left(1+\e'\sum_{n=1}^{l}\fr1n \right)(1-\g_0\e') \nn\\
 \G(1+x)& \simeq & 1-\g_0 x
 \eea
where $\g_0$ is Euler's constant and using a formula derived in
the appendix, (\ref{4g}),  we get
 \bea
 \fr{C_2(-1)^l}{C_1-(-1)^lC_2}
 &\simeq & \left(\fr{Q_1Q_5(w^2-\l^2)}{4R^4}\right)^{{l+1}}
   \g^{2(l+1)} \fr{\xi_0}{\e[l!(l+1)!]^2} \nn\\
   && \left[\;\; \fr{1}{2}
  \left( \fr{\G'(1+\a+\b+\fr{l}{2})}{\G(1+\a+\b+\fr{l}{2})}
 +\fr{\G'(1+\b-\a+\fr{l}{2})}{\G(1+\b-\a+\fr{l}{2})}
 \right.\right. \nn\\
 &&\left.\left.+\fr{\G'(\a+\b-\fr{l}{2})}{\G(\a+\b-\fr{l}{2})}
  +\fr{\G'(\b-\a-\fr{l}{2})}{\G(\b-\a-\fr{l}{2})}
  \right)
         +\pi\;\fr{\cos \pi(\b-\a+\fr{l}2)}{\sin
\pi(\b-\a+\fr{l}2)}  \right.\nn\\
 && \left. -\sum_{n=1}^{l}\fr1n +2\g_0
 - \fr{\xi_1}{\xi_0}  \;\; \right] \label{ccomplex}
 \eea
where the prime on $\G$ represents a derivative with respect to
the argument and $\xi \equiv \xi_0+\e'\;\xi_1+\cdots$. The first
two coefficients, $\xi_0$ and $\xi_1$, are derived in the
appendix. They are given by
 \bea\label{xi0}
 \xi_0 &=& \fr{\G(1+\a+\b+\fr{l}2)}{\G(\a+\b-\fr{l}2)}
        \fr{\G(1+\b-\a+\fr{l}2)}{\G(\b-\a-\fr{l}2)} \nn\\
  \fr{\xi_1}{\xi_0} &=& -\fr12\fr{\G'(1+\a+\b+\fr{l}{2})}{\G(1+\a+\b+\fr{l}{2})}
 -\fr12\fr{\G'(1+\b-\a+\fr{l}{2})}{\G(1+\b-\a+\fr{l}{2})} \nn\\
 &&+\fr12\fr{\G'(\a+\b-\fr{l}{2})}{\G(\a+\b-\fr{l}{2})}
 +\fr12\fr{\G'(\b-\a-\fr{l}{2})}{\G(\b-\a-\fr{l}{2})}
 \eea
Substituting (\ref{xi0}) into (\ref{ccomplex}), one gets
 \bea
 \fr{C_2(-1)^l}{C_1-(-1)^lC_2}
&\simeq & \left(\fr{Q_1Q_5(w^2-\l^2)}{4R^4}\right)^{{l+1}}
   \g^{2(l+1)} \fr{\xi_0}{\e[l!(l+1)!]^2} \nn\\
&& \left[\;\;
   \fr{\G'(1+\a+\b+\fr{l}{2})}{\G(1+\a+\b+\fr{l}{2})}
 +\fr{\G'(1+\b-\a+\fr{l}{2})}{\G(1+\b-\a+\fr{l}{2})}
 \right. \nn\\
 && \left.+\pi\;\fr{\cos \pi(\b-\a+\fr{l}2)}
 {\sin\pi(\b-\a+\fr{l}2)}
   -\sum_{n=1}^{l}\fr1n +2\g_0
   \;\; \right]
 \eea
This, combined with (\ref{csimple}), allows one to re-write
(\ref{R}),
 \bea
 R&=&e^{-i\pi\e}-2\pi i\;
 \left(\fr{Q_1Q_5(w^2-\l^2)}{4R^4}\right)^{{l+1}}
   \g^{2(l+1)} \fr{\xi_0}{[l!(l+1)!]^2} \nn\\
&& \left[\;\;
   \fr{\G'(1+\a+\b+\fr{l}{2})}{\G(1+\a+\b+\fr{l}{2})}
 +\fr{\G'(1+\b-\a+\fr{l}{2})}{\G(1+\b-\a+\fr{l}{2})}
 \right. \nn\\
 && \left.+\pi\;\fr{\cos \pi(\b-\a+\fr{l}2)}
 {\sin\pi(\b-\a+\fr{l}2)}
   -\sum_{n=1}^{l}\fr1n +2\g_0
   \;\; \right] \label{R2}
 \eea
Depending on the number of trips that a quantum makes inside the
tube, each of them has a different phase. To discover the phase,
we rewrite the $\fr{\cos}{\sin}$-term as
 \bea
 \fr{\cos \pi(\b-\a+\fr{l}2)}{\sin\pi(\b-\a+\fr{l}2)}
 &=& -i\;\fr{1+e^{2\pi(\b-\a+{l}/2)}}{1-e^{2\pi(\b-\a+{l}/2)}}
       \nn\\
 &=&-i \left(1+2\sum_{n=1}^{\infty}\;e^{2\pi i n(\b-\a+{l}/2)}\right)
 \eea
where in the second identity, we have employed the formal
expansion that was used in \cite{Lunin:2001dt}. With this $R$
becomes,
 \bea
 R&=& \left[e^{-i\pi\e}-2\pi i\;
 \left(\fr{Q_1Q_5(w^2-\l^2)}{4R^4}\right)^{{l+1}}
   \g^{2(l+1)} \fr{\xi_0}{[l!(l+1)!]^2} \right. \nn\\
&& \left.\times\left(\;\;
   \fr{\G'(1+\a+\b+\fr{l}{2})}{\G(1+\a+\b+\fr{l}{2})}
 +\fr{\G'(1+\b-\a+\fr{l}{2})}{\G(1+\b-\a+\fr{l}{2})}
 -i\pi  -\sum_{n=1}^{l}\fr1n +2\g_0
   \;\; \right)\right] \nn\\
 &&-4\pi^2 \;
 \left(\fr{Q_1Q_5(w^2-\l^2)}{4R^4}\right)^{{l+1}}
   \g^{2(l+1)} \fr{\xi_0}{[l!(l+1)!]^2}\sum_{n=1}^{\infty}\;e^{2\pi i n(\b-\a+{l}/2)}
 \eea
From the last term which contains the exponential factor, one can
read off the time delay between two adjacent quanta that emerge
from the tube,
 \be \label{timedelay}
 \Delta t=2\pi\frac{d}{d\tilde\omega}(\beta-\alpha)=
 \pi\frac{R}{\gamma}=\pi\frac{\sqrt{Q_1Q_5}}{a}.
 \ee
This is precisely the same as the result in \cite{Lunin:2001dt}:
neither the momentum on the compact direction nor the generic
value of $\g$ affects the time delay. However, both of them do
affect the probability for a quantum to go into the throat. The
probability is given as a square of the coefficient of the
exponential factor,
 \bea\label{probability}
 P&=&\left[4\pi^2 \;
 \left(\fr{Q_1Q_5(w^2-\l^2)}{4R^4}\right)^{{l+1}}
   \fr{\g^{2(l+1)}}{[l!(l+1)!]^2} \;\xi_0 \right]^2
    \\
&=&\left[4\pi^2 \;
 \left(\fr{Q_1Q_5(w^2-\l^2)}{4R^4}\right)^{{l+1}}
   \fr{\g^{2(l+1)}}{[l!(l+1)!]^2}\fr{\G(1+\a+\b+\fr{l}2)}{\G(\a+\b-\fr{l}2)}
        \fr{\G(1+\b-\a+\fr{l}2)}{\G(\b-\a-\fr{l}2)} \right]^2 \nn
\eea
where in the second line we have used the first equation of
(\ref{xi0}). This reduces, as one can easily verify, to the result
of \cite{ms,Lunin:2001dt} when $\l=0$ and $\g\rightarrow 0$, but
differs from it generically.

\bigskip

In this note, we have considered the geometry given by
(\ref{MaldToCompare}) and a quantum placed inside. We have
computed two quantities: the probability for a quantum to enter
the throat, (\ref{timedelay}), and the time interval for it to
round-trip the throat once, (\ref{probability}). We conclude with
a few remarks. The geometry of (\ref{MaldToCompare}) is such that
in the throat region it is AdS whereas it becomes flat
asymptotically. The dual CFT, which is an orbifold CFT, describes
the AdS region. As shown above, the time delay turns out to be
the same as the result of orbifold CFT that was obtained through
an `effective string' picture \cite{Lunin:2001jy}. This
strengthens the validity of the use of the orbifold CFT for the
physics inside the throat region. Meanwhile, the flat region plays
a role, as one might have expected, for the probability. In the
CFT analysis, the flat region is accounted by terms that break
conformal symmetry. In the leading order the `non-CFT' terms give
results that agree with the gravity calculation. Since we now
have more general results in the gravity side, it should be
possible to match them with CFT computation by including more
detailed and higher order terms that arise from expanding a DBI
action. Our results should be useful for that purpose.

 \vspace{.5in}
 \ni{\bf \large Acknowledgements \\}
 I thank S. D. Mathur for suggesting this problem. I am also very grateful
 to O. Lunin for valuable discussions. This work is
 in part supported by DOE grant DE-FG02-91ER-40690.

\newpage

\appendix
\section*{Appendices}

\section{Useful Formulae}

\renewcommand{\theequation}{A.\arabic{equation}}
\setcounter{equation}{0}

 Define $\xi \equiv \xi_0+\e' \xi_1$. In the first order in $\e'$,
 \bea
 \xi &=& \fr{\G(q+l+1)\G(q-2\a+l+1)}{\G(q)\G(q-2\a)} \nn\\
     &= &
     \fr{\G(1+\a+\b+\fr{l}2-\fr{\e'}2)\G(1+\b-\a+\fr{l}2-\fr{\e'}2)}
     {\G(\a+\b-\fr{l}2-\fr{\e'}2)\G(\b-\a-\fr{l}2-\fr{\e'}2)}\nn\\
     &\simeq & \fr{\G(1+\a+\b+\fr{l}2)-\fr{\e'}2\G(1+\a+\b+\fr{l}2)}
     {\G(\a+\b-\fr{l}2)-\fr{\e'}2\G'(\a+\b-\fr{l}2)} \nn\\
     &&\times\fr{\G(1+\b-\a+\fr{l}2)-\fr{\e'}2\G'(1+\b-\a+\fr{l}2)}
     {\G(\b-\a-\fr{l}2)-\fr{\e'}2\G'(\b-\a-\fr{l}2)}\nn\\
 &\simeq & \xi_0 \left(1+\fr{\e'}{2}
  \left( -\fr{\G'(1+\a+\b+\fr{l}{2})}{\G(1+\a+\b+\fr{l}{2})}
 -\fr{\G'(1+\b-\a+\fr{l}{2})}{\G(1+\b-\a+\fr{l}{2})}
 \right.\right. \nn\\
 && \left.\left.+\fr{\G'(\a+\b-\fr{l}{2})}{\G(\a+\b-\fr{l}{2})}
 +\fr{\G'(\b-\a-\fr{l}{2})}{\G(\b-\a-\fr{l}{2})}
  \right)
         \right)
 \eea
where the prime on $\G$'s represents a derivative with respect to
the argument. From this one reads off
 \bea\label{g1overg0}
 \xi_0 &=& \fr{\G(1+\a+\b+\fr{l}2)}{\G(\a+\b-\fr{l}2)}
        \fr{\G(1+\b-\a+\fr{l}2)}{\G(\b-\a-\fr{l}2)} \nn\\
  \fr{\xi_1}{\xi_0} &=& -\fr12\fr{\G'(1+\a+\b+\fr{l}{2})}{\G(1+\a+\b+\fr{l}{2})}
 -\fr12\fr{\G'(1+\b-\a+\fr{l}{2})}{\G(1+\b-\a+\fr{l}{2})} \nn\\
 &&+\fr12\fr{\G'(\a+\b-\fr{l}{2})}{\G(\a+\b-\fr{l}{2})}
 +\fr12\fr{\G'(\b-\a-\fr{l}{2})}{\G(\b-\a-\fr{l}{2})}
 \eea
The $\fr{\G\G}{\G\G}$-term that appears e.g., in (\ref{c2}) can be
approximated as follows,
 \bea\label{gggg}
 && \fr{\G(\fr12+\a+\b+\fr{\n'}{2})
   \G(\fr12+\a-\b+\fr{\n'}{2})}
   {\G(\fr12+\a+\b-\fr{\n'}{2})\G(\fr12+\a-\b-\fr{\n'}{2})} \nn\\
 = && \fr{\G(\fr12+\a+\b+\fr{\n'}{2})
   \G(\fr12+\b-\a+\fr{\n'}{2})}
   {\G(\fr12+\a+\b-\fr{\n'}{2})\G(\fr12+\b-\a-\fr{\n'}{2})}\;\;
 \fr{\sin\pi (\a-\b+\fr{1-\n'}{2})}{\sin\pi (\a-\b+\fr{1+\n'}{2})}
 \nn\\
  \simeq &&  \fr{\G(1+\a+\b+\fr{l}{2}+\fr{\e'}{2})
   \G(1+\b-\a+\fr{l}{2}+\fr{\e'}{2})}
   {\G(\a+\b-\fr{l}{2}-\fr{\e'}{2})\G(\b-\a-\fr{l}{2}-\fr{\e'}{2})}\nn\\
 &&\left( 1+\pi\e'\;\fr{\cos \pi(\b-\a+\fr{l}2)}{\sin \pi(\b-\a+\fr{l}2)}\right)
  (-1)^{l+1}
 \eea
where in the first equality we have used
 \be
 \Gamma(x)=\frac{\pi}{\sin\pi x}\frac{1}{\Gamma(1-x)},
 \ee
and in the second equality a trigonometric identity. In the first
order of $\e'$ the four $\G$-piece that appears in the third line
of (\ref{gggg}) can be written as
 \bea
&&\fr{\G(1+\a+\b+\fr{l}{2}+\fr{\e'}{2})
   \G(1+\b-\a+\fr{l}{2}+\fr{\e'}{2})}
   {\G(\a+\b-\fr{l}{2}-\fr{\e'}{2})\G(\b-\a-\fr{l}{2}-\fr{\e'}{2})}\nn\\
 \simeq && \fr{\left(\G(1+\a+\b+\fr{l}{2})+\fr{\e'}{2} \G'(1+\a+\b+\fr{l}{2})\right)
                    }
      {\left(\G(\a+\b-\fr{l}{2})-\fr{\e'}{2} \G'(\a+\b-\fr{l}{2})\right)
           } \nn\\
 && \times\frac{\left(\G(1+\b-\a+\fr{l}{2})+\fr{\e'}{2}
  \G'(1+\b-\a+\fr{l}{2})\right)}
  {\left( \G(\b-\a-\fr{l}{2})-\fr{\e'}{2} \G'(\b-\a-\fr{l}{2})\right)}
 \nn\\
 \simeq && \fr{\G(1+\a+\b+\fr{l}{2})\G(1+\b-\a+\fr{l}{2})}
 {\G(\a+\b-\fr{l}{2})\G(\b-\a-\fr{l}{2})}
 \left(1+\fr{\e'}{2}\fr{\G'(1+\a+\b+\fr{l}{2})}{\G(1+\a+\b+\fr{l}{2})} \right)
 \nn\\
 &&\left(1+\fr{\e'}{2}\fr{\G'(1+\b-\a+\fr{l}{2})}{\G(1+\b-\a+\fr{l}{2})}
 \right)
 \left(1+\fr{\e'}{2}\fr{\G'(\a+\b-\fr{l}{2})}{\G(\a+\b-\fr{l}{2})}
 \right)
 \left(1+\fr{\e'}{2}\fr{\G'(\b-\a-\fr{l}{2})}{\G(\b-\a-\fr{l}{2})} \right)
  \nn\\
\simeq &&  \xi_0 \left(1+\fr{\e'}{2}
  \left( \fr{\G'(1+\a+\b+\fr{l}{2})}{\G(1+\a+\b+\fr{l}{2})}
 +\fr{\G'(1+\b-\a+\fr{l}{2})}{\G(1+\b-\a+\fr{l}{2})}
 \right.\right. \nn\\
 && \left.\left. +\fr{\G'(\a+\b-\fr{l}{2})}{\G(\a+\b-\fr{l}{2})}
 +\fr{\G'(\b-\a-\fr{l}{2})}{\G(\b-\a-\fr{l}{2})}
  \right)
         \right)
 \eea
Therefore
 \bea\label{4g}
 && \fr{\G(\fr12+\a+\b+\fr{\n'}{2})
   \G(\fr12+\a-\b+\fr{\n'}{2})}
   {\G(\fr12+\a+\b-\fr{\n'}{2})\G(\fr12+\a-\b-\fr{\n'}{2})} \nn\\
\simeq &&  \xi_0 \left[1+\fr{\e'}{2}
  \left( \fr{\G'(1+\a+\b+\fr{l}{2})}{\G(1+\a+\b+\fr{l}{2})}
 +\fr{\G'(1+\b-\a+\fr{l}{2})}{\G(1+\b-\a+\fr{l}{2})}
 \right.\right. \\
 && \left.\left. +\fr{\G'(\a+\b-\fr{l}{2})}{\G(\a+\b-\fr{l}{2})}
 +\fr{\G'(\b-\a-\fr{l}{2})}{\G(\b-\a-\fr{l}{2})}
  \right)
         \right]
         \left( 1+\pi\e'\;\fr{\cos \pi(\b-\a+\fr{l}2)}{\sin
         \pi(\b-\a+\fr{l}2)}\right)
  (-1)^{l+1} \nn
 \eea


\newpage

\begin{thebibliography}{99}



\bibitem{sv}
A. Strominger and C. Vafa, { Phys. Lett.} {\bf B379}, 99 (1996),
hep-th/9601029.


\bibitem{cm}
C. Callan and J. Maldacena, { Nucl. Phys.} {\bf B472}, 591
(1996), hep--th/9602043.

\bibitem{dm}
S.R. Das and S.D. Mathur, { Nucl. Phys.} {\bf B478}, 561 (1996),
hep--th/9606185.

\bibitem{ms}
J.~Maldacena and A.~Strominger, Phys. Rev.  {\bf D55} (1997) 861,
hep--th/9609026.

\bibitem{Mathur:2002ie}
S.~D.~Mathur,
Int.\ J.\ Mod.\ Phys.\ D {\bf 11}, 1537 (2002)
[arXiv:hep-th/0205192].


\bibitem{Lunin:2001jy}
O.~Lunin and S.~D.~Mathur,
Nucl.\ Phys.\ B {\bf 623}, 342 (2002) [arXiv:hep-th/0109154].





\bibitem{mm}
V.~Balasubramanian, J.~de Boer, E.~Keski-Vakkuri and S.~F.~Ross,
hep-th/0011217; J.~Maldacena and L.~Maoz,
hep-th/0012025.

\bibitem{Lunin:2002bj}
O.~Lunin, S.~D.~Mathur and A.~Saxena,
Nucl.\ Phys.\ B {\bf 655}, 185 (2003) [arXiv:hep-th/0211292].

\bibitem{Lunin:2001dt}
O.~Lunin and S.~D.~Mathur,
Nucl.\ Phys.\ B {\bf 615}, 285 (2001) [arXiv:hep-th/0107113].










\bibitem{cvetic}
M.~Cvetic and D.~Youm,
Nucl.\ Phys.\ B {\bf 476}, 118 (1996), hep-th/9603100.


\bibitem{larsentwo}
M.~Cvetic and F.~Larsen,
Phys.\ Rev.\ D {\bf 56}, 4994 (1997), hep-th/9705192.
%











\end{thebibliography}
\end{document}